\documentclass[aps,prc,reprint,superscriptaddress]{revtex4-2}

\usepackage{graphicx}
\usepackage{dcolumn}
\usepackage{bm}
\usepackage{longtable}
\usepackage{amsmath,amssymb}
\usepackage{mathrsfs}
\usepackage{flushend}
\usepackage{microtype}
\usepackage{array} 
\usepackage{booktabs}
\usepackage{newtxtext,newtxmath}
\usepackage{hyperref} 
\hypersetup{
    colorlinks=true,
    linkcolor=blue,
    filecolor=blue,      
    urlcolor=blue,
    citecolor=blue,
}

\begin{document}

\title{Fourier shape parametrization in covariant density functional theory for nuclear fission}

\author{Zeyu Li}
\affiliation{China Nuclear Data Center, China Institute of Atomic Energy, Beijing 102413, China}
\affiliation{School of Physical Science and Technology, Southwest University, Chongqing 400715, China}

\author{Yang Su}
\affiliation{China Nuclear Data Center, China Institute of Atomic Energy, Beijing 102413, China}

\author{Lile Liu}
\affiliation{China Nuclear Data Center, China Institute of Atomic Energy, Beijing 102413, China}

\author{Yongjing Chen}
\email{ahchenyj@126.com}
\affiliation{China Nuclear Data Center, China Institute of Atomic Energy, Beijing 102413, China}

\author{Zhipan Li}
\email{zpliphy@swu.edu.cn}
\affiliation{School of Physical Science and Technology, Southwest University, Chongqing 400715, China}

\date{\today}

\begin{abstract}
\fontsize{10pt}{12pt}\selectfont
We implement the Fourier shape parametrization within the point-coupling covariant density functional theory to construct the collective space, potential energy surface (PES), and mass tensor, which serve as inputs for the time-dependent generator coordinate method to simulate the fission dynamics. Taking \(^{226}\)Th as a benchmark, we demonstrate the superiority of Fourier shape parametrization over conventional spherical harmonic parametrization: it significantly enhances the convergence of higher-order collective shape parameters by efficiently characterizing extreme nuclear deformations. Consequently, the new framework generates more reasonable elongated configurations, particularly for the scission configurations, and significantly improves the description of charge distribution near the symmetric fission peak. Moreover, the Fourier shape parametrization provides a smooth and well-defined three-dimensional (3D) PES with minimal correlations between degrees of freedom, enabling high-precision 3D dynamical simulations of fission.
\end{abstract}

\maketitle
\fontsize{10.8pt}{12.4pt}\selectfont

\section{Introduction}\label{Introduction}

Nuclear fission presents a unique example of non-equilibrium large-amplitude collective motion where all nucleons participate with complex correlation effects, making the microscopic description of fission one of the most complex problems in low-energy nuclear physics \cite{Bender2020JPG,Schmidt2018Review,Schunck2022PPNP}.
Since the discovery of nuclear fission, various theories have been put forward and made great progress \cite{Schunck2016RPP,Hans2012Book,Schunck2022PPNP, Bertsch2015Review,Andreyev2018Review,Schmidt2018Review,Bulgac2020Review}. 
Based on the work of Bohr and Wheeler \cite{Bohr1939}, the early theories for fission introduced a set of deformation parameters into the liquid drop model to construct multi-dimensional potential energy surfaces (PESs) to describe the relationship between nuclear deformation and energy,  which gives a simple explanation and clear picture of nuclear fission.
In subsequent studies, shell corrections and pair correlations have been added to the liquid drop model, which is called the macroscopic-microscopic (MM) approach \cite{BRACK1972RMP,Nix1972ARNS}. The MM approach has a series of versions characterized by different parametrizations of nuclear shape of the liquid drop and different phenomenological nuclear potentials, such as the five-dimensional finite-range liquid-drop model (FRLDM) \cite{Moller2001Nature,Moller2009PRC,Ichikawa2012PRC,CSchmitt2020PRC}, macroscopic-microscopic Woods-Saxon model \cite{Jachimowicz2012PRC,Jachimowicz2013PRC}, the macroscopic–microscopic Lublin–Strasbourg drop (LSD) model in the three-quadratic-surface parametrization \cite{Wang2019CTP,Zhu2020CTP}, the LSD in Fourier shape parametrization \cite{Schmitt2017PRC,Pomorski2023PRC}, two-center shell model \cite{Liu2019PRC,LiuLL2022PRC} and so on. Based on a large number of parameters, the MM approach has greatly optimized the description of atomic nuclei. However, due to the parameter dependence of the results, the explanation of the microscopic mechanism of fission still eludes us.

Modern microscopic approaches to fission are predominantly grounded in nuclear density functional theory (DFT), which provides a self-consistent framework for addressing both static and dynamic aspects of fission \cite{Schunck2016RPP,Schmidt2018RPP,Simenel2018PPNP,Bender2020JPG,Verriere2020FP,Schunck2022PPNP,PeiJC2023CSBC,GuoL2024EPJA,ZhouSG2023IJMPE,Goddard2015PRC,Godbey2024PRC,WangXB2023PRC}. Within the adiabatic approximation of DFT, the total energies and wave functions along the fission path are determined by minimizing the energy density functional of the nucleus under specific constraints and assumed symmetries. Fission observables, such as fission yields and total kinetic energy distributions, are then derived by evolving the collective wave packet on the microscopic PES using methods like the time-dependent generator coordinate method (TDGCM) \cite{Berger1984NPA,Goutte2005PRC,Regnier2016PRC,Regnier2016CPC,Zdeb2017PRC,Regnier2019PRC,Younes2012LLNL,TaoH2017PRC,ZhaoJ2019PRC,ZhaoJ2019PRC2,ZhaoJ2021PRC,ZhaoJ2022PRC,ZhaoJ2022PRC2,chen_energy_2022,chen_microscopic_2023,Chen_pair_2023,Schunck_oddA_2023,LiZY2022PRC,LiZY2024TCHO,RegnierD2019PRC,RegnierD2020FIP}. In this approach, the dynamics of the fissioning system depend critically on microscopic inputs, such as the PES and collective inertia, which are functions of a few collective coordinates.

However, fully microscopic and nonadiabatic time-dependent DFT studies have revealed that fission involves the excitation of many collective degrees of freedom \cite{Bulgac2016PRL,Libo2024FP,RenZX2022PRL}. Given the limitations of current computing technology, the calculation of potential energy surfaces is constrained to a finite number of dimensions. Consequently, a key challenge is to define a collective space that effectively captures nuclear fission configurations within these finite-dimensional constraints.

\begin{figure}[htbp]
  \centering
  \includegraphics[width=9cm]{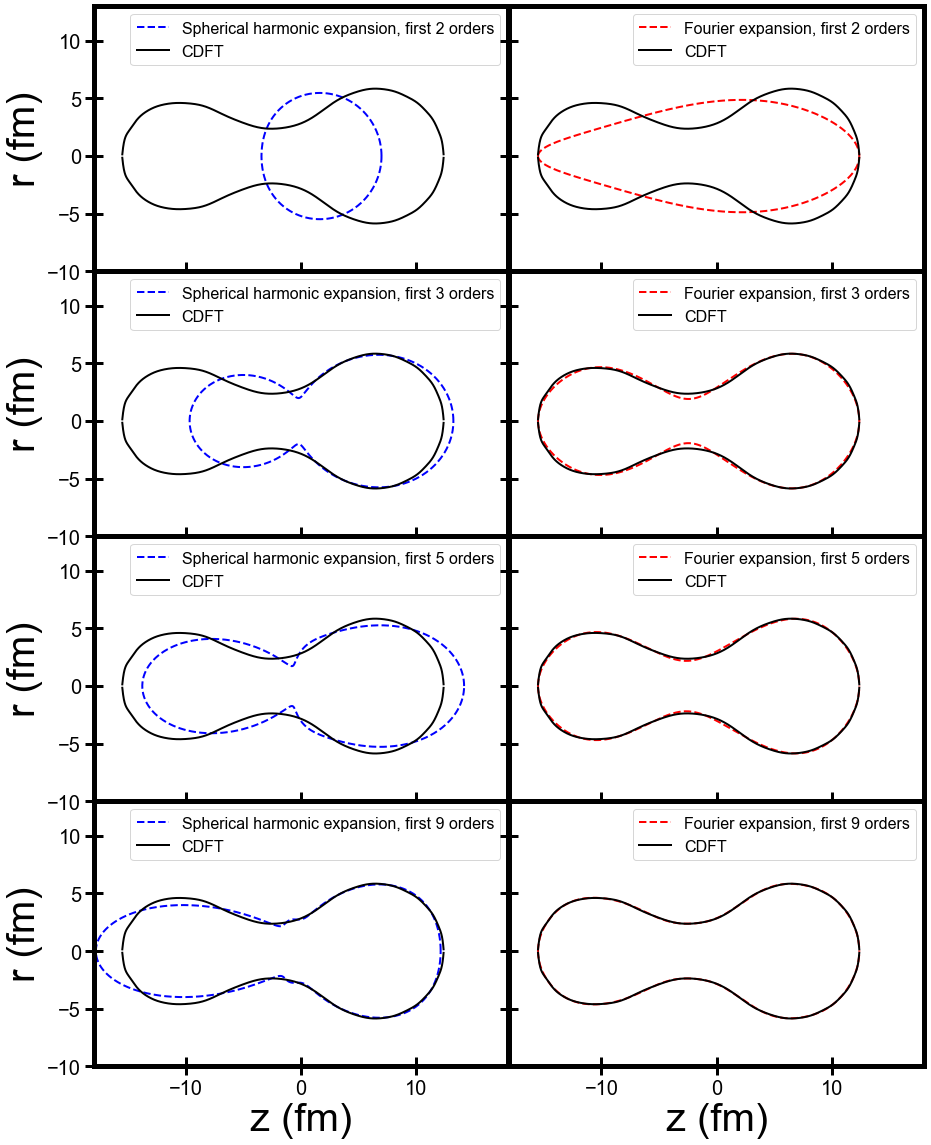}
  \caption{Convergence of the spherical harmonic shape parametrization (left, blue line) and Fourier shape parametrization (right, red line) for a typical nuclear shape on the fission path of $^{226}$Th derived from CDFT calculations (black solid line).}
  \label{expand}
\end{figure}

To identify an optimal collective deformation space, it is essential to develop methods that describe nuclear shapes with minimal variables while preserving accuracy. Existing studies often use spherical harmonic (SH) expansion within the microscopic DFT framework to parameterize nuclear shapes. To evaluate the precision of this expansion, the left panels of Fig. \ref{expand} illustrate the convergence process of SH expansion for a typical nuclear shape along the fission path in covariant DFT (CDFT) calculations. Notably, the convergence remains poor even up to the 9th order, highlighting the limitations of SH parametrization for nuclear fission \cite{Schunck2016RPP,Schunck2022PPNP}.

In contrast, various MM approaches have proposed alternative shape parametrizations, such as the Nilsson perturbed-spheroid parameterization \cite{MOLLER20161}, Funny Hills shape description \cite{RevModPhys.44.320}, the two-center shell model \cite{maruhn_asymmetrie_1972}, Los Alamos parameterization \cite{NIX19651}, Cassini ovals \cite{PASHKEVICH1971275}, and Fourier shape parameterization \cite{FourierPRC.95.034612,Pomorski2015APPB}. These methods have demonstrated advantages in describing nuclear shapes for fission within finite dimensions. In particular, the Fourier shape parameterization, which describes nuclear shapes through a Fourier series expansion, exhibits rapid convergence for shapes along the fission path, as shown in the right panels of Fig. \ref{expand}. Remarkably, even the first three terms of the Fourier expansion provide an accurate representation of the nuclear shape.

In this work, we implement the Fourier shape parametrization within the point-coupling CDFT framework. Using $^{226}$Th as a benchmark, we perform illustrative calculations for the PES, scission configurations, and induced fission dynamics. The advantages of the Fourier shape parametrization are highlighted through a comparative analysis with the results obtained from the SH shape parametrization. Section \ref{Theory} outlines the theoretical framework, while Section \ref{Benchmark} provides a detailed comparison and discussion of the PES and fragment yield distributions calculated using CDFT and TDGCM. Finally, Section \ref{Summary} summarizes the key findings and offers an outlook for future research.

\section{Theoretical framework}\label{Theory}
\begin{figure}[htbp]
  \centering
  \includegraphics[width=8.5cm]{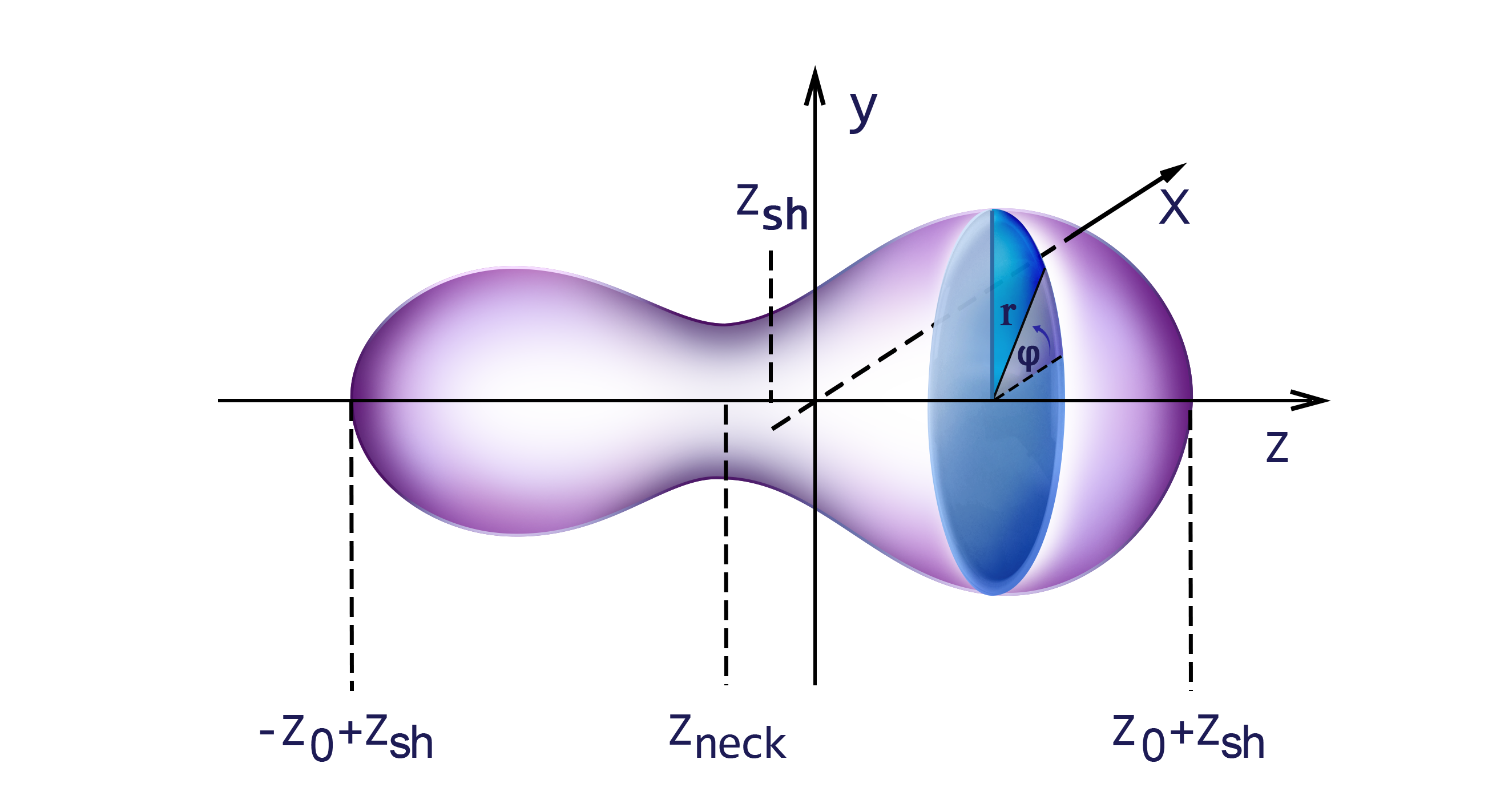}
  \caption{Schematic illustration of a fissioning nucleus in a cylindrical coordinate system, where \( z_{\rm sh} \) and \( z_0 \) denote the center and half-length of the nucleus along the \( z \)-axis, respectively.}
  \label{profile}
\end{figure}

Nuclear fission is a slow, large-amplitude collective motion that can be described as the time evolution of a collective wave function governed by specific collective degrees of freedom. In this work, we employ a recently proposed Fourier shape parametrization to construct the collective space. For an axially symmetric nuclear shape, the profile function is expanded as a Fourier series \cite{FourierPRC.95.034612}:
\begin{gather}
  \begin{split}
    r_s^2(z) = &R_0^2\sum_{n=1}^{\infty}\left[a_{2n}\text{cos}(\frac{(2n-1)\pi}{2}\frac{z-z_{\rm sh}}{z_0}) \right.\\
                                &\ \ \ \ \ \ \ \ \ \ \ \left.+ a_{2n+1}\text{sin}(\frac{2n\pi}{2}\frac{z-z_{\rm sh}}{z_0})\right].
  \end{split}
  \label{eq:rs}
\end{gather}
Here, $r_s(z)$ represents the distance from a surface point at coordinate $z$ to the symmetric axis ($z$-axis in Fig. \ref{profile}), and $R_0$ is the radius of a spherical nucleus with the same volume, calculated as $R_0 = 1.2 A^{1/3}$ fm. $z_{\rm sh}$ and $z_0$ denote the geometric center position relative to the center of mass (anchored at the origin of coordinate system) and the half-length of the nucleus along the $z$-axis, respectively, which are determined by \cite{FourierPRC.95.034612}:
\begin{gather}
  \begin{split}
    z_0    &= \frac{\pi R_0}{3\sum_n (-1)^{n-1} a_{2n}/(2n-1)} \\
    z_{\rm sh} &= \frac{3z_0^2}{2\pi R_0}  \sum_n (-1)^n \frac{a_{2n+1}}{n}
  \end{split}
  \label{eq:z0zsh}
\end{gather}

Equations (\ref{eq:rs}, \ref{eq:z0zsh}) define the Fourier shape parameters for a uniform liquid drop with a sharp surface. However, realistic nuclei exhibit non-uniform nucleon density distributions with diffused surface, such as those obtained from self-consistent DFT calculations. In such cases, the Fourier shape parameters must be determined using corresponding operators. Starting from nucleon number conservation:
\begin{gather}
  A = \rho_0\int \pi r_s^2(z) {\rm d}z = \int \rho_{\rm DFT}(z, r, \phi) r{\rm d}r{\rm d}\phi {\rm d}z
\end{gather}
where $\rho_0=\frac{A}{\frac{4}{3}\pi R_0^3}$ and $\rho_{\rm DFT}(z, r, \phi)$ is the density from DFT calculations, the Fourier shape parameters are derived as:
\begin{gather}
  \begin{split}
    a_{2n}   &= \int  \frac{4R_0 \text{cos}(\frac{(2n-1)\pi(z-z_{\rm sh})}{2z_0})}{3Az_0}  \rho_{\rm DFT}(z,r,\phi) r{\rm d}r{\rm d}\phi {\rm d}z \\
    a_{2n+1} &= \int  \frac{4R_0 \text{sin}(\frac{2n\pi(z-z_{\rm sh})}{2z_0})}{3Az_0}      \rho_{\rm DFT}(z,r,\phi) r{\rm d}r{\rm d}\phi {\rm d}z \\
  \end{split}
\end{gather} 
with their corresponding operators:
\begin{gather}
  \begin{split}
    \hat{A}_{2n}  &= \frac{4R_0}{3Az_0} \text{cos}(\frac{(2n-1)\pi(z-z_{\rm sh})}{2z_0})\\
    \hat{A}_{2n+1}&= \frac{4R_0}{3Az_0} \text{sin}(\frac{2n\pi(z-z_{\rm sh})}{2z_0})
  \end{split}
\end{gather}

The entire map of the energy surface in multi-dimensional collective space for fission is obtained by imposing constraints on a number of Fourier shape parameters $a_k$
\begin{equation}
  \langle E_{\rm tot}\rangle+\sum\limits_{k=2, 3, \cdots}C_k(\langle\hat A_k\rangle-a_k)^2
\end{equation}
where $\langle E_{\rm tot}\rangle$ is the total energy of DFT and $C_k$  are the corresponding stiffness constants. 

Since the Fourier shape parameters $a_k$ are not as intuitively related to the shape of a nucleus \cite{FourierPRC.95.034612,Pomorski2015APPB}, we define a new collective space $q_k$:
\begin{gather}
  \begin{split}
    q_2 &= \frac{a_2^{(0)}}{a_2} - \frac{a_2}{a_2^{(0)}} \\
    q_3 &= a_3 \\
    q_4 &= a_4 + \sqrt{(q_2/9)^2+(a_4^{(0)})^2}.
  \end{split}
\end{gather}
Here, $a_2^{(0)}$ and $a_4^{(0)}$ represent the corresponding values of Fourier shape parameters when the nucleus is spherical. The new collective variables $q_k$ are more directly connected to nuclear shapes: $q_2$ describes overall elongation, $q_3$ reflects mass asymmetry, and $q_4$ influences neck thickness.

When the configurations for the full collective space are obtained under the constrained calculations, we can finally determine the collective PES by subtracting the zero-point motion energy (ZPE), e.g. the vibrational ZPE
\begin{equation}
  \label{eq:Vcoll}
  V(q_2, q_3, \cdots)=E_{\rm tot}-\Delta E_{\rm vib}.
  \end{equation}
The ZPE can be calculated in the cranking approximation \cite{Girod1979NPA} and the expression for vibrational ZPE reads
\begin{equation}
  \label{ZPE-vib}
  \Delta E_{\rm vib} = \frac{1}{4} \textnormal{Tr}\left[\mathcal{M}_{(3)}^{-1}\mathcal{M}_{(2)}  \right]\;,
  \end{equation}
with
\begin{equation}
  \label{masspar-M}
  \mathcal{M}_{(n),kl}=\sum_{i,j}
   {\frac{\left\langle i\right|\hat{A}_{k}\left| j\right\rangle
   \left\langle j\right|\hat{A}_{l}\left| i\right\rangle}
   {(E_i+E_j)^n}\left(u_i v_j+ v_i u_j \right)^2}\;.
  \end{equation}
Here, $E_i$ and $v_i$ are the quasiparticle energies and occupation probabilities, respectively. The summation is over the proton and neutron single-particle states in the canonical basis. 

For the dynamical simulation, one also needs to calculate the mass tensor in the perturbative cranking approximation \cite{Girod1979NPA}. Firstly, the mass tensor in the collective space $(a_2, a_3, \cdots)$ is given by:
\begin{equation}
\label{eq:BB}
B_{kl}(a_2, a_3, \cdots)=\hbar^2 \left[\mathcal{M}_{(1)}^{-1} \mathcal{M}_{(3)} \mathcal{M}_{(1)}^{-1}\right]_{kl}.
\end{equation}
and is then transformed to the new collective space $(q_2, q_3,\cdots)$:
\begin{gather}
    B_{ij}(q_2, q_3, \cdots) = \sum_{kl} \frac{\partial a_k}{\partial q_i} \frac{\partial a_l}{\partial q_j} B_{kl}(a_2, a_3, \cdots)
\end{gather}

Using the PES and mass tensor as inputs, low-energy fission dynamics can be simulated by a time-dependent equation derived from the TDGCM in the Gaussian overlap approximation
\begin{align}
  & i \hbar \frac{\partial}{\partial t} g(q_{2}, q_{3}, t) \notag \\
  & = \left[-\frac{\hbar^{2}}{2} \sum_{ij} \frac{\partial}{\partial q_{i}} B^{-1}_{ij}\left(q_{2}, q_{3}\right) \frac{\partial}{\partial q_{j}}+ V(q_{2}, q_{3})\right] \notag \\
  &\ \ \ \times g(q_{2}, q_{3}, t) \label{eq:Schrodinger_like}
\end{align}
where $g(q_2, q_3, t)$ is a complex wave function, which contains all the information about the dynamics of the system. At present, we solve the time-dependent equation in the two-dimensional collective space $(q_2, q_3)$ using the FELIX code package \cite{Regnier2018CPC}. The yield of fission fragments with mass $A_H$ can be obtained by integrating the probability current $\mathbf{J}$ that runs through the scission line.
\begin{gather}
  Y(A_H) \propto \sum_{\xi \in \mathcal{A}} \lim _{t \rightarrow+\infty} \int_{t=0}^{t} d t \int_{\left(q_{2}, q_{3}\right) \in \xi} \mathbf{J}\left(q_{2}, q_{3}, t\right) \cdot d \mathbf{S}
  \label{yields_eq}
\end{gather}
where $\mathcal{A}$ is the set of all elements $\xi$ belonging to the scission line such that the heavy fragment has mass $A_H$. More details on TDGCM could be found in \cite{LiZY2024TCHO,Regnier2016CPC,Regnier2018CPC}.

\section{Illustrative calculations for $^{226}$Th}\label{Benchmark}
\subsection{Numerical details}
In this section, we present illustrative calculations of the static and dynamic fission properties of \(^{226}\)Th. Specifically, we compare the PESs, scission configurations, and charge yield distributions calculated using the TDGCM based on CDFT in two shape spaces: Fourier shape space \((q_2, q_3)\) and SH shape space \((\beta_2, \beta_3)\). For the CDFT calculations, the energy density functional PC-PK1 \cite{Zhao2010PRC} is used to describe the effective interaction in the particle-hole channel, while a \(\delta\) force is employed in the particle-particle channel. The strength parameters of the \(\delta\) force, \(V_n = 360\) MeV fm\(^3\) and \(V_p = 378\) MeV fm\(^3\), are determined by reproducing the empirical pairing gaps using a five-point formula \cite{Bender2000EPJA}. The Dirac equation is solved using a two-center harmonic oscillator basis with a major shell cutoff of \(N_f = 20\) \cite{LiZY2024TCHO}.

Large-scale deformation-constrained CDFT calculations are performed to generate the PESs, scission lines, and mass tensors in the \((q_2, q_3)\) plane. The collective variables \(q_2\) and \(q_3\) are varied over the ranges \(-0.6 \leq q_2 \leq 3.0\) (with a step size \(\Delta q_2 = 0.02\)) and \(0.0 \leq q_3 \leq 0.4\) (with a step size \(\Delta q_3 = 0.005\)). In the collective space, scission is characterized by a discontinuity between the prescission and postscission configurations. Following our previous work \cite{TaoH2017PRC}, the prescission domain is defined by a neck nucleon number \(q_N \geq 3\), with the boundary of this domain identified as the scission line.

For the induced fission dynamics, the TDGCM is employed to model the time evolution of the fissioning nucleus using a time step \(\delta t = 5 \times 10^{-4}\) zs. To account for the escape of the collective wave packet outside the calculation region, an imaginary absorption potential is introduced with an absorption rate \(r = 30 \times 10^{22}\) s\(^{-1}\) and an absorption band width \(w = 3.0\).

\subsection{Results for $^{226}$Th}
\begin{figure}[htbp]
  \centering
  \includegraphics[width=9cm]{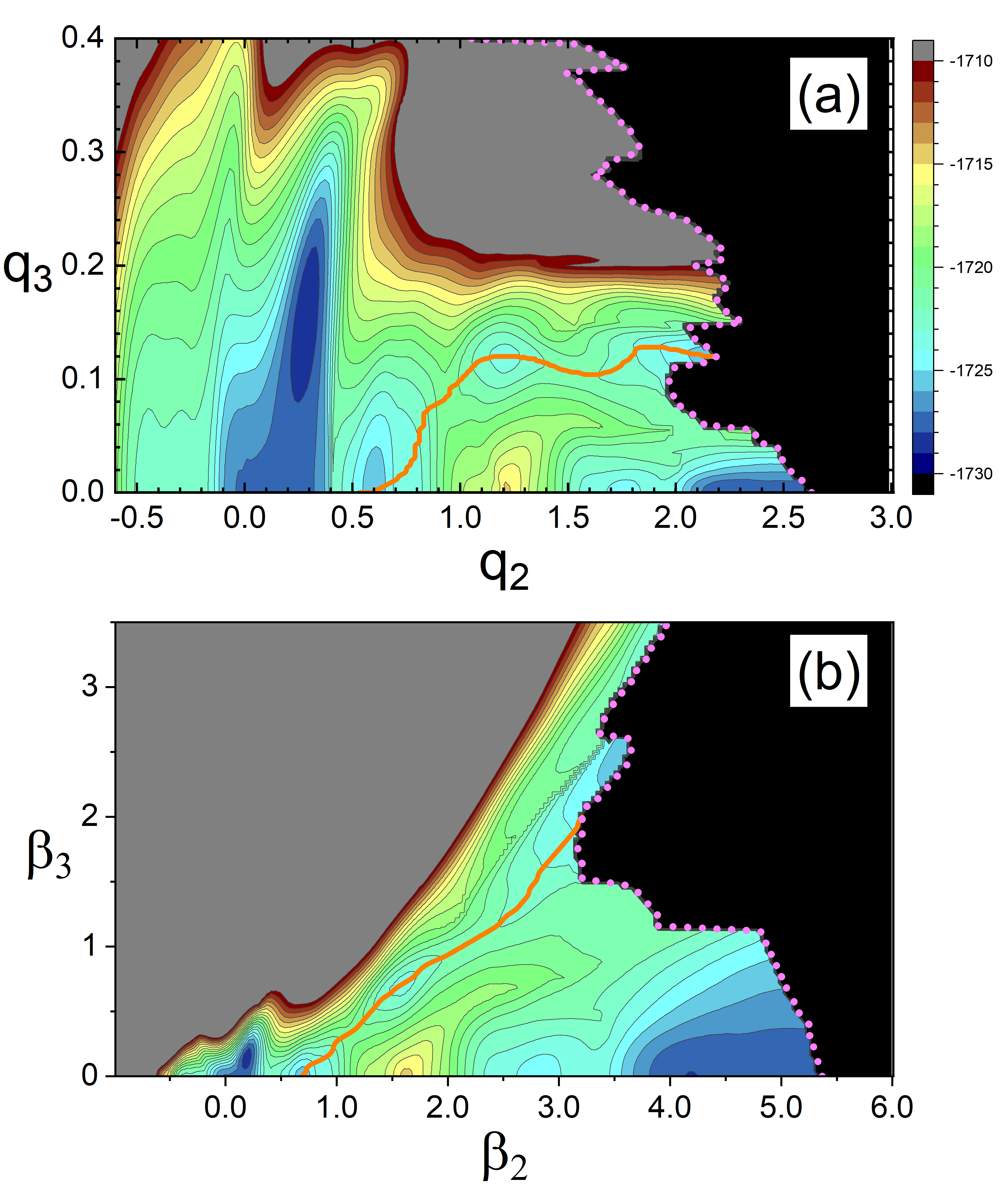}
  \caption{ (Color online) The potential energy surfaces of $^{226}\text{Th}$ in the Fourier shape space $(q_2, q_3)$ (panel a) and spherical harmonic shape space $(\beta_2, \beta_3)$ (panel b) calculated from constrained CDFT with PC-PK1 density functional. The optimal asymmetric fission path and scission line are denoted by the orange solid lines and pink dotted lines, respectively.}
  \label{PES}
\end{figure}

\begin{figure}[htbp]
  \centering
  \includegraphics[width=8.5cm]{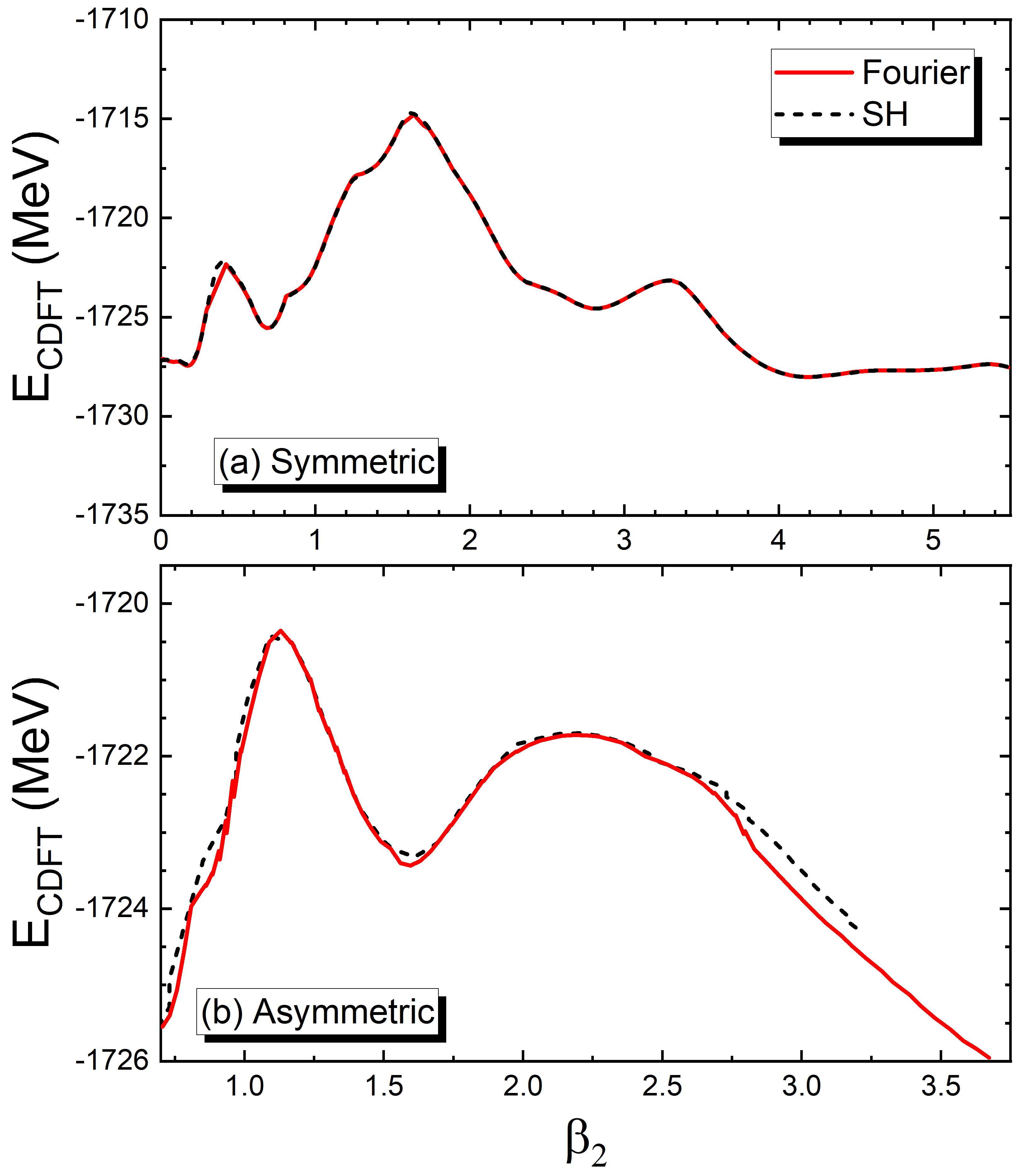}
  \caption{ (Color online) The potential energy curves along the symmetric (a) and optimal asymmetric (b) fission paths of $^{226}$Th calculated using constrained CDFT with the Fourier and SH shape parametrizations, respectively.}
  \label{path}
\end{figure}

\begin{figure}[htbp]
  \centering
  \includegraphics[width=8.5cm]{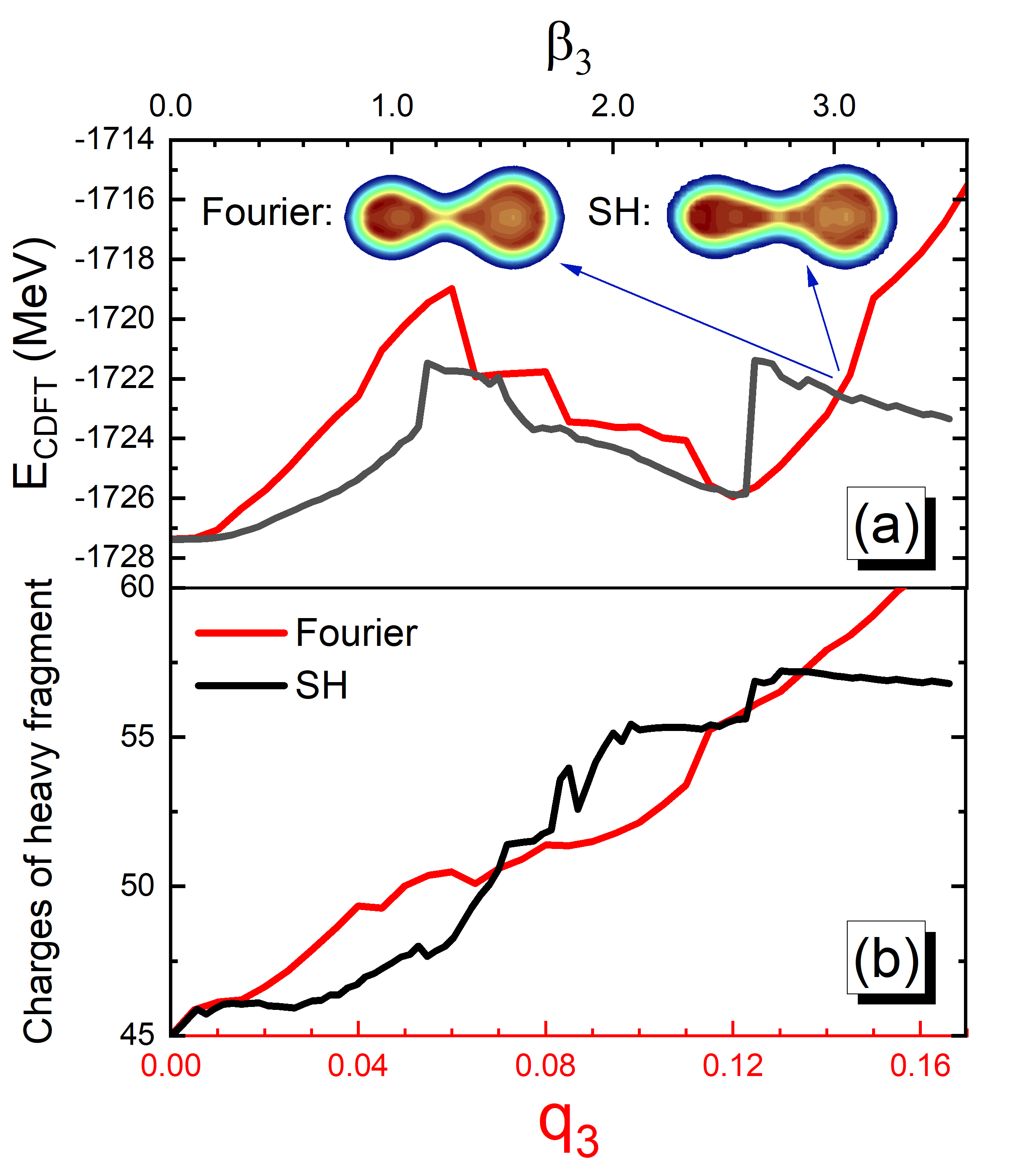}
  \caption{(Color online) (a) Scission configuration energies and (b) heavy fragment charge numbers as functions of mass-asymmetric collective variables $q_3$ (Fourier) and $\beta_3$ (SH). Density distributions for two shape parametrizations are also shown in panel (a) for comparison.}
  \label{scission}
\end{figure}

Figure \ref{PES} displays the PESs of $^{226}\text{Th}$ in the Fourier shape space $(q_2, q_3)$ and SH shape space $(\beta_2, \beta_3)$ calculated from constrained CDFT. The optimal asymmetric fission path and scission line are denoted by the orange solid lines and pink dotted lines, respectively. The topography of PES in the $(q_2, q_3)$ plane closely resembles that obtained from the macroscopic-microscopic approach \cite{FourierPRC.95.034612}, validating the Fourier shape parametrization based on CDFT. The equilibrium state locates at $(q_2, q_3)=(0.29,0.16)$, which corresponds to the minimum at $(\beta_2, \beta_3)=(0.19, 0.14)$ in the SH shape space, and both are quite soft with respect to the mass asymmetry coordinates. Two PESs exhibit coexisting symmetric and asymmetric fission valleys separated by a ridge. Notably, the asymmetric fission valley in the Fourier shape space (denoted by the solid line) remains at \(q_3 \approx 0.13\) beyond the second fission barrier, indicating that the fission fragments are essentially formed and maintained until scission. In contrast, the asymmetric fission valley in the SH shape space shows a strong correlation between \(\beta_2\) and \(\beta_3\). A detailed comparison of the fission valleys and scission lines for the two shape parametrizations is provided in Figs. \ref{path} and \ref{scission}, respectively.

Figure \ref{path} shows the potential energy curves (PECs) for the symmetric fission paths (\(q_3 = 0\) or \(\beta_3 = 0\)) and the optimal asymmetric fission paths (orange solid lines in Fig. \ref{PES}) as functions of the quadrupole shape parameter \(\beta_2\). The quadrupole shape parameter \(\beta_2\) for the Fourier shape parametrization is directly calculated from the nucleon density distribution derived from constrained CDFT. For the symmetric fission path, the PECs nearly overlap, indicating identical configurations along the path. For the asymmetric fission path, while the two PECs are generally consistent, slight differences are observed in the barrier and valley regions, with a maximum deviation of $\sim 0.2$ MeV (see Table \ref{Barrier Heights}). Notably, the asymmetric fission path in the Fourier shape space is lower and more extended than that in the SH shape space for \(\beta_2 > 2.5\). This is attributed to the Fourier shape parametrization's superior accuracy in representing highly elongated configurations due to its rapid convergence, as demonstrated in Table \ref{tab:Expand}, which lists the Fourier and SH shape parameters of the third fission barrier along the optimal asymmetric fission path. In the Fourier shape parametrization, the first three shape parameters have relatively large values, while higher-order terms decrease by one or two orders of magnitude, highlighting the fast convergence of the Fourier parametrization for large elongated configurations. In contrast, the SH shape parametrization shows no convergence up to the 7th order and even exhibits an divergent behavior.

\begin{table}[h!]
  \begin{center}
    \caption{Fission barrier heights (in MeV) with respect to the corresponding equilibrium state of $^{226}$Th for the Fourier and SH shape parametrizations.}
  \label{Barrier Heights}
  \begin{tabular}{p{1.3cm}p{1.3cm}p{1.3cm}p{1.3cm}p{1.3cm}p{1.3cm}}
    \toprule
    \toprule
             & $B_I$  & $B_{II}^{Sym}$& $B_{III}^{Sym}$&$B_{II}^{Asym}$&$B_{III}^{Asym}$ \\ 
    \midrule
      Fourier&6.59&14.14& 5.77& 8.57&7.20\\ 
    \midrule
      SH &6.66& 14.12& 5.65& 8.38&  7.10\\ 
    \bottomrule
    \bottomrule
  \end{tabular}  
  \end{center}
\end{table}

\begin{table}[h!]
  \begin{center}
  \caption{The Fourier and SH shape parameters of the third fission barrier along the optimal asymmetric fission path calculated using constrained CDFT with their corresponding shape parametrizations. For clarity, the original Fourier shape parameters $(a_2, a_3, \cdots)$ are presented here to better illustrate their convergence behavior. \label{tab:Expand}}
  \begin{tabular}{p{1.1cm}p{1.1cm}p{1.1cm}p{1.1cm}p{1.1cm}p{1.1cm}p{1.1cm}}
    \toprule
    \toprule
      $n$ &~~~2~~~&~~~3~~~&~~~4~~~&~~~5~~~&~~~6~~~&~~~7~~~ \\ 
    \midrule
      $a_n$&0.4865&0.1050& -0.2607& -0.0165&-0.0096&-0.0131 \\ 
    \midrule
      $\beta_n$ &2.2329 & -1.0275 & 3.5623 & -3.1828 &  6.2295&-7.2577 \\ 
    \bottomrule
    \bottomrule
  \end{tabular}  
  \end{center}
\end{table}

Fig. \ref{scission} presents the scission configuration energies and heavy fragment charge numbers calculated using Fourier and SH shape parametrizations as a function of their respective mass-asymmetric shape parameters. The curves exhibit distinct minima at $q_3=0$ and $\beta_3=0$, corresponding to symmetric fission valley configurations. With increasing mass asymmetry, both energy curves initially rise before descending to secondary minima at $q_3=0.12$ and $\beta_3=2.56$, respectively, marking the asymmetric fission valley configurations (cf. Fig. \ref{PES}). Notably, both parametrizations yield consistent scission configurations at these minima, showing identical energy values ($E \sim -1726$ MeV) and heavy fragment charges ($Z_H \sim 56$). For highly mass-asymmetric scission configurations, two calculations exhibit quite different trends: energies in the SH shape space increase sharply before gradually decreasing, whereas those in the Fourier shape space rise steadily. Importantly, density distributions in this region differ significantly, especially for the light fragment.

Regarding the heavy fragment charge numbers ($Z_H$) shown in Fig. \ref{scission}(b), the SH shape parametrization results display several plateaus and abrupt jumps. This leads to an overcounting of certain fragment configurations, such as $Z_H \approx 55$ and 57, while missing others, particularly for heavier fragments with $Z_H > 57$, as experimentally observed \cite{Schmidt2000NPA}. This issue is also evident in other DFT calculations employing SH shape parametrization \cite{Regnier2016PRC,ZhaoJ2019PRC,chen_microscopic_2023}. In contrast, the Fourier shape parametrization successfully addresses these artifacts by establishing a smooth, nearly linear dependence between $Z_H$ and  $q_3$. As a result, it can capture a wider range of charge numbers, especially for heavier fragments with $Z_H > 57$. A similar linear relationship is observed between the heavy fragment mass number $A_H$ and $q_3$. 

\begin{figure}[htbp]
  \centering
  \includegraphics[width=8.5cm]{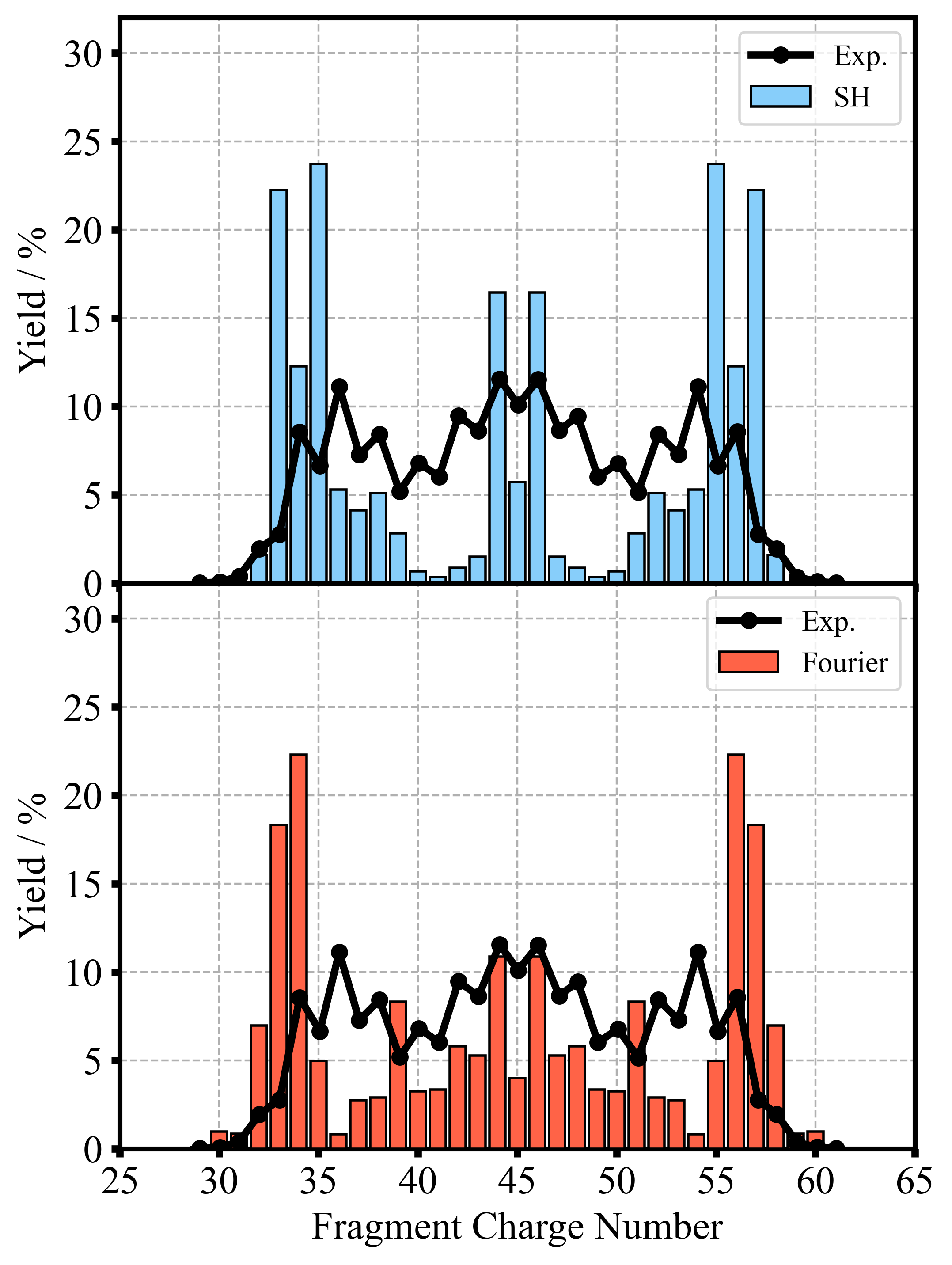}
  \caption{ (Color online) The fission fragment charge distributions calculated based on ($\beta_2,\beta_3$) plane (top) and ($q_2,q_3$) plane (bottom). The experimental charge yields are also shown for comparison \cite{Schmidt2000NPA}.}
  \label{yields}
\end{figure}

Using the PESs, mass tensors, and scission configurations as inputs, we simulate the photoinduced fission dynamics of \(^{226}\)Th within the TDGCM framework. Following the procedure of our previous work \cite{TaoH2017PRC}, the initial state is prepared by boosting the collective ground state along the \(\beta_2\) (or \(q_2\)) degree of freedom to achieve a target excitation energy of 11 MeV \cite{Schmidt2000NPA}. Figure \ref{yields} presents the preneutron emission charge yields for \(^{226}\)Th calculated using both the SH and Fourier shape parametrizations. Both calculations successfully reproduce the experimental trend, capturing the coexistence of symmetric and asymmetric fission peaks. Notably, the Fourier shape parametrization provides a more accurate description of the charge distribution in the range \(38 \leq Z_f \leq 52\), as it better accommodates elongated configurations, particularly near scission.

Despite significant improvements in describing the symmetric peak (Fig. \ref{yields}), discrepancies remain for the asymmetric peaks compared to experimental data. These deviations likely stem from the inherent limitations of two-dimensional collective space calculations, underscoring the necessity of extending the simulation to three dimensions.

\begin{figure}[htbp]
  \centering
  \includegraphics[width=9cm]{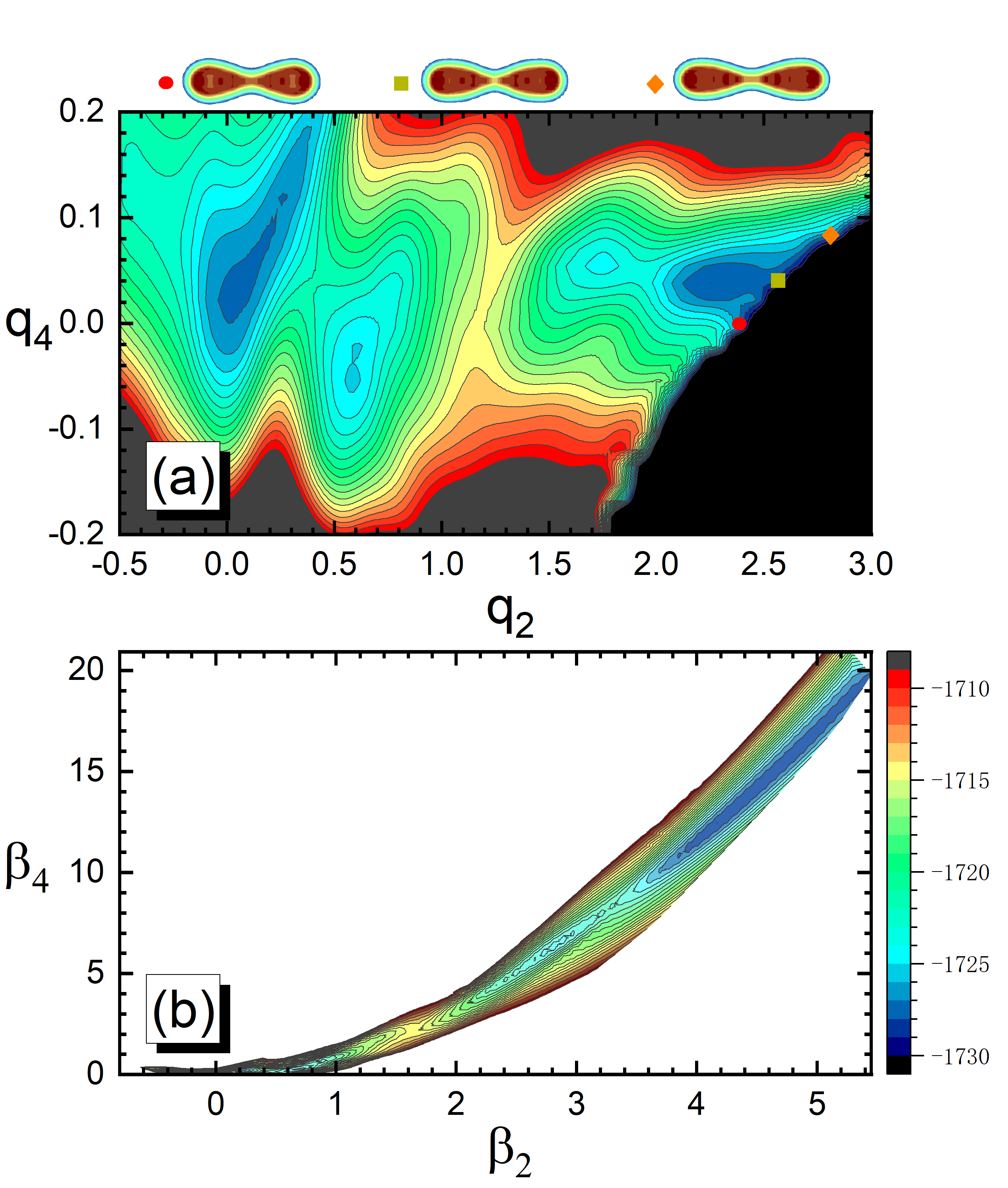}
  \caption{(Color online) The potential energy surfaces of $^{226}\text{Th}$ in the Fourier shape space $(q_2, q_4)$ with $q_3 = 0$ (panel a) and SH shape space $(\beta_2, \beta_4)$ with $\beta_3 = 0$ (panel b).}
  \label{q24PES}
\end{figure}

As a preliminary investigation, Fig. \ref{q24PES}(a) shows the PES of \(^{226}\)Th in the \((q_2, q_4)\) plane with \(q_3 = 0\). The PES reveals a distinct structure: an equilibrium state at \((q_2, q_4) = (0.05, 0.02)\) with a soft behavior along \(q_4\), a fission valley within \(q_4 \in [-0.1, 0.1]\), and three saddle points along the fission path. For the scission configurations, the PES is rather soft across an extended region from \((q_2, q_4)\approx(2.3, 0.0)\) to \(\approx (3.0, 0.1)\), resulting in significant fluctuations in fragment deformations and distances (see the density distributions for selected scission configurations in Fig. \ref{q24PES}(a)). These fluctuations further influence fragment angular momenta, deformation energies, total kinetic energies, and other observables. 

For comparison, Fig. \ref{q24PES}(b) depicts the PES in the SH shape space \((\beta_2, \beta_4)\) with \(\beta_3 = 0\). The strong correlation between \(\beta_4\) and \(\beta_2\) leads to a narrow PES, obscuring detailed structural features, especially near scission, and complicating dynamical simulations using methods such as the TDGCM. In our previous work, this limitation was partially addressed by introducing the neck nucleon number \(q_N\) as a third collective variable in place of \(\beta_4\) \cite{Zhou_2023}. However, the global definition of \(q_N\) remains problematic, posing challenges for multi-dimensional dynamical simulations of nuclear fission. In contrast, the newly implemented Fourier shape parameterization appears to resolve these issues, enabling robust multi-dimensional dynamical simulations of nuclear fission.

\section{Summary and outlook\label{Summary}}

In this study, we have derived the Fourier shape operators and implemented the Fourier shape parametrization within the framework of point-coupling covariant density functional theory (CDFT) to construct the collective space, potential energy surfaces (PESs), and mass tensors for nuclear fission. The fission dynamics is simulated using the time-dependent generator coordinate method (TDGCM) with the Gaussian overlap approximation. 

Taking \(^{226}\)Th as a benchmark, we have systematically validated the advantages of the Fourier shape parametrization by comparing it with macroscopic-microscopic calculations using Fourier shape parametrization and microscopic calculations based on the conventional spherical harmonic (SH) shape parametrization. The key findings are summarized as follows:

1. \textbf{Validity of Fourier shape parametrization within microscopic CDFT}: The PES in the Fourier shape space \((q_2, q_3)\) closely aligns with that obtained from the macroscopic-microscopic approach, confirming the validity of the Fourier shape parametrization within CDFT. Notably, the asymmetric fission valley remains stable at \(q_3 \approx 0.13\) beyond the second fission barrier, indicating that fission fragments are preformed and persist until scission. In contrast, the SH shape parametrization exhibits a strong correlation between \(\beta_2\) and \(\beta_3\), obscuring the detailed structure of the fission valley.

2. \textbf{Superior Convergence and Accuracy}: The Fourier shape parametrization demonstrates rapid convergence, with the first three shape parameters dominating the description of highly elongated configurations, while higher-order terms diminish by \(1\sim 2\) orders of magnitude. This results in a smooth, nearly linear relationship between fragment charge (mass) and the Fourier mass-asymmetric shape parameter \(q_3\), enabling a broader and more accurate representation of scission configurations across a wide range of \(Z_H\) (\(A_H\)) values. In contrast, the SH shape parametrization shows no convergence up to the 7th order and even exhibits divergent behavior. 

3. \textbf{Improved Dynamical Calculations}: The above advancements significantly enhance the accuracy of TDGCM simulations, which utilize the PES and mass tensor as inputs. Specifically, the Fourier shape parametrization provides a more precise description of the charge distribution in the range \(38 \leq Z_f \leq 52\).

Despite these improvements, discrepancies remain in the description of asymmetric fission peaks, likely due to the inherent limitations of two-dimensional collective space calculations. To address this, we have recently performed a fully three-dimensional (3D) calculation for the compound nucleus \(^{236}\)U, incorporating constraints on axial quadrupole and octupole deformations \((\beta_2, \beta_3)\) as well as the neck nucleon number \(q_N\) \cite{Zhou_2023}. This work underscores the importance of 3D calculations in elucidating quantum fluctuations and considering more fission modes. However, the global definition of \(q_N\) remains problematic, posing challenges for 3D dynamical simulations. In contrast, the Fourier shape parametrization inherently avoids such limitations, providing a smooth and well-defined 3D PES with minimal correlations between degrees of freedom. These features highlight the feasibility of extending the TDGCM framework to 3D using Fourier shape parametrization, a task currently underway.

\begin{acknowledgments}
This work was partly supported by the National Natural Science Foundation of China (Grants No. 12375126 and No. 12275081),  the Fundamental Research Funds for the Central Universities, and the Continuous-Support Basic Scientific Research Project.
\end{acknowledgments}

\bibliography{FourierCDFT.bib,review_fission.bib,NuclearFission_Theory_Classics.bib}
\end{document}